\documentstyle[epsfig,a4,12pt]{article}

\begin{document}

\newcommand \be  {\begin{equation}}
\newcommand \bea {\begin{eqnarray} \nonumber }
\newcommand \ee  {\end{equation}}
\newcommand \eea {\end{eqnarray}}
\newcommand{\siml}{\stackrel{<}{\sim}}
\newcommand{\simg}{\stackrel{>}{\sim}}


\begin{titlepage}
    
\title{\bf BUBBLES, CRASHES AND INTERMITTENCY IN AGENT BASED MARKET MODELS}

\author{Irene Giardina$^{1,2}$ and Jean-Philippe Bouchaud$^{3,4}$ \\
\small{$^1$ Service de Physique Th\'eorique,
 Centre d'\'etudes de Saclay,}\\  
\small{Orme des Merisiers, 
91191 Gif-sur-Yvette Cedex, France}\\
\small{$^2$ Dipartimento di Fisica, Universit\`a di Roma La Sapienza,}\\
\small{Piazzale 
Aldo Moro 2, 00185 Roma, Italy} \\
\small{$^3$ Service de Physique de l'\'Etat Condens\'e,
 Centre d'\'etudes de Saclay,}\\  
\small{Orme des Merisiers, 
91191 Gif-sur-Yvette Cedex, France}\\
\small{$^4$ Science \& Finance, CFM, 109-111 rue Victor-Hugo, 92532 France}}

\maketitle

\begin{abstract}
We define and study a rather complex market model, 
inspired from the Santa Fe artificial market and the Minority Game. Agents have 
different strategies among which they can choose, according to their relative profitability,
with the possibility of not participating to the market. The price is updated 
according to the excess demand, and the wealth of the agents is properly accounted for. 
Only two parameters play a significant role: one describes the impact of trading on the 
price, and the other describes the propensity of agents to be trend following or contrarian. 
We observe three different regimes, depending on the
value of these two parameters: an oscillating phase with bubbles and crashes, an intermittent 
phase and a stable `rational' market phase. The statistics of price changes in 
the intermittent phase resembles that of real price changes, with small linear 
correlations, fat tails and long range volatility clustering. We discuss how 
the time dependence of these two parameters spontaneously drives the system in 
the intermittent region.
We analyze quantitatively the temporal correlation of activity in the intermittent
phase, and show that the `random time strategy shift' mechanism that we proposed 
earlier allows one to understand the observed long ranged correlations. Other mechanisms leading 
to long ranged correlations are also reviewed. We discuss several other issues, 
such as the formation of bubbles and 
crashes, the influence of transaction costs and the distribution of agents wealth. 
\end{abstract}
\end{titlepage}

\section{Introduction}

It is now well known that the statistics of price changes in financial
markets exhibit interesting `stylized facts', which are to some extent
universal, i.e. independent of the type of market (stocks, currencies, 
interest rates, etc.) and of the epoch \cite{Guillaume,MS,Book,farmerrev}. 
Price changes are in a good
approximation uncorrelated beyond a time scale of the order of tens of
minutes (on liquid markets). Their distribution is strongly non Gaussian:
they can be characterized by Pareto (power-law) tails with an exponent
in the range $3-5$. Another striking feature is the intermittent nature of the
fluctuations: localized outbursts of the volatility, i.e.
the amplitude of the price fluctuations (averaged over a given time interval), 
can be identified. 
This fact, known as {\it volatility clustering} 
\cite{volfluct1,volfluct2,MS,Book}, 
can be analyzed more quantitatively: the temporal correlation function
of the daily volatility $\sigma_t$ can be 
fitted by an inverse power of the lag $\tau$, with a rather small exponent in
the range $0.1 - 0.3$ \cite{volfluct2,PCB,stanley,muzy,MRW}. This suggests that there is no characteristic time scale
for volatility fluctuations: outbursts of market activity can persist for 
short times (a few hours), but also for much longer times, months or even years. 
The slow decay
of the volatility correlation function leads to a multifractal-like 
behaviour of price changes \cite{Ghash,Mandelbrot,Fisher,MRW,BMP}, and has important consequences for option pricing.
Other stylized facts have been reported, such as the leverage effect that
leads to skewed distribution of price changes \cite{leverage}, or the apparent increase
of inter-stock correlations in volatile periods.

It is now very clear to many that these features are very difficult to explain within the 
traditional framework of `rational expectations', where all agents
share the same information, have an infinite computation power and act in a 
perfectly rational way (see e.g the clear discussion 
in the introduction of refs. \cite{Hommes,Chiarella} and in \cite{Schiller}). 
Another route, much less formalized and still very much in an exploratory stage, is followed by an increasing number of academics. 
The aim is to assume as little as possible about agents preferences and 
abilities, and to explore {\it generic} classes of models, with the hope of finding
some {\it plausible} mechanisms that reproduce at least part of the stylized facts
recalled above.  In this endeavor, one 
should not be constrained by preexisting prejudices or established frameworks.  
The `grand unification' of different mechanisms which would lead in fine to
a logically consistent and simultaneous 
understanding of all the empirical facts is deferred to 
later times. Similarly, it is premature to ask for rigorous proofs, but leave
space for hand waving arguments and numerical simulations.

In this paper, we report \cite{PhysicaA} the results of an artificial market that bears some
similarities with many previous attempts 
\cite{BG,SantaFe,Arthur,Levy,MG,Bak,BC,Farmer,Iori,Stauffer,MG2,Lux1,Lux2,Farmer2,Johnson,Johnson2,Marsili,Raberto,Andersen-Sornette}. 
Although the detailed 
behaviour of our artificial market depends on the value 
of the different parameters entering the model, only a few market typologies
are identified, and some qualitative features (such as the long range 
volatility correlations) are robust to parameter changes, at least in some
regions of parameter space. We explain in 
particular that the general mechanism proposed in \cite{QF} is indeed responsible
for volatility clustering in our model. We also discuss, in this context, 
other agent based 
models proposed so far in which this phenomenon has been observed. We identify 
several other interesting features (appearance of bubbles and crashes, 
influence of the price fixing procedure and of wealth constraints, of transaction costs, etc.) that may be of some relevance to real markets. 

In our model, we have not included herding, or imitation effects. Each agent
acts independently of other agents. The correlations between their actions is entirely
mediated by the price history itself. Direct herding might also be important to account for the phenomenology of real markets (see \cite{orlean,CB,Stauffer,IOP} and references therein). We leave to a future study the extension of the present model to account for these herding effects.

The aftermath of our study (and of all other similar studies) is the following paradox:
in order to get a `good looking' price chart, one has to tune quite a bit the important 
parameters of the model. What are the mechanisms tuning the parameters in 
real markets to make them look all alike ? There must be some generic self-organization
mechanisms responsible for this selection. We discuss in section \ref{efficiency}, 
in view of our results, what 
could be the `evolutionary' driving forces relevant for this fundamental issue.

\section{Set up of the model}

\subsection{Basic ingredients}

In line with the original idea of the Santa Fe artificial market \cite{SantaFe,Arthur}, which was later
simplified and popularized as the Minority Game ({\sc mg}) \cite{MG,MG1,MG2,MG3,mulet}, 
in our model agents do not follow a rational expectation paradigm but rather act inductively, 
adapting their
behaviour to their past experience. As in the {\sc mg} each agent has a
certain fixed number of strategies, each of which converts some {\it information} into a {\it decision}.
We will assume a world where there are only two tradable assets: a stock, with
fluctuating price, and a bond, yielding a certain (known) risk free rate $\rho$. The information 
on which the agents decide their action will be the past history of the price itself.
The decision is to buy stocks (converting bonds in cash), 
to sell stocks, or to be inactive (i.e. to hold bonds). Each strategy 
is given a score, which is updated according to its performance.
The strategy played at time $t$ by a given agent is the one, among those available to him, 
which would have best performed in a recent past. 
We take proper account of the wealth balance of each agent, 
and proper market clearing (i.e matching supply and demand) is enforced. 

\subsection{Notations and definitions}

Each agent $i$, $i \in \{1,...,N\}$ has $S-1$ active strategies plus an inactive
one. He owns, at time $t$, a number $\phi_i(t)$ of stocks and $B_i(t)$ of bonds. 
The price of the stock is $X(t)$, and therefore the total wealth of agent $i$ is $B_i(t)+
\phi_i(t)X(t)$. The dynamics of the model, between $t$ and $t+1$, is defined by the following set of rules:

\begin{itemize}
\vskip 0.3cm
\item{Information:}
\vskip 0.3cm
We assume that all agents rely on the {\it same} information ${\mathcal I}_t$, given by the $m$ 
last steps of the past history of the return time series ($m$ is for `memory'). 
We choose the information to be qualitative and to only depend on the {\it sign} of the previous price changes: 
\be
{\mathcal I}_t=\{\chi(t-m),\cdots ,\chi(t-1)\} \qquad \chi(t)= {\rm {sign}}
\left[\log (\frac{X(t)}{X(t-1)})-\rho\right].
\label{history}
\ee
In this sense, our traders are `chartists' on short time scales, and take
their decision based on the past pattern of price changes. (We will add below the
possibility for the agents not to follow their systematic strategies, and act as
`fundamental' traders if the price is too far from a reference value, or even act
at random). 

A comment on the value of the time scale `$1$' is in order here. Clearly, different agents
observe the price time series on different time scales, from several minutes for intra day
traders to months for long term pension funds. Here, we assume for simplicity that all
agents coarse-grain the price time series using the same clock, and consider as meaningful
price variations on -- say -- a day or a week. All the parameters below were chosen such
as one time step roughly corresponds to a week of trading. One interesting outcome of our model 
is that even if all agents have the same intrinsic clock, a broad range of time scales is
spontaneously generated.
\vskip 0.3cm
\item{Strategies:}
\vskip 0.3cm
Each agent $i$ is endowed with a certain number $S$ of fixed strategies, 
that convert information ${\mathcal I}_t$
into decision $\epsilon_i({\mathcal I}_t)=\pm 1, 0$ (buy, sell, inactive). For example, a `trend following' 
strategy could be to choose $\epsilon_i=1$ as soon as there is a majority of $+$'s in the
signal ${\mathcal I}_t$. Another example is the inactive strategy, 
for which $\epsilon_i \equiv 0$, $\forall {\mathcal I}$.

Each agent is given the possibility to remain inactive, i.e. has an inactive strategy.
The $S-1$ other strategies are chosen at random in the space of all strategies (there are $2^{2^m}$
of them), in order to model heterogeneity in the agents capabilities.
One can however give a bias to this random choice, and favor `trend following' 
or `contrarian' strategies. This is done by defining the `magnetization' $M \in [-1,1]$ of the
string $\chi(1),\cdots ,\chi(m)$, defined as:
\be
M= \frac{1}{m} \sum_{j=1}^m \chi(j),
\ee
and choose the corresponding decision to be $\epsilon=+1$ with probability $(1+P M)/2$ and 
$\epsilon=-1$ with probability $(1-P M)/2$. The parameter $P \in [-1,1]$ can be called the `polarization'
of the strategies. If $P=1$, trend following strategies are favored, whereas $P=-1$ 
corresponds to contrarian strategies. The choice $P=0$ is means no bias in the strategy space.
\vskip 0.3cm
\item{Decision and buy orders:}
\vskip 0.3cm 
Knowing the strategy used by  agent $i$ and the information ${\mathcal I}_t$ allows one to 
compute the decision $\epsilon_i(t)$ of each agent. Depending on the value of 
$\epsilon_i(t)$, the agent buys/sells a quantity $q_i(t)$ proportional to his current 
belongings. More precisely, we set:
\bea
q_i(t)= g \frac{B_i(t)}{X(t)} \qquad &{\rm for}& \ \epsilon_i(t) = +1 \\
q_i(t)= - g \phi_i(t) \qquad &{\rm for}& \ \epsilon_i(t) = -1 \nonumber  \\  
q_i(t)= 0 \qquad &{\rm for}& \ \epsilon_i(t) = 0.
\eea
This means that we consider `prudent' investors who change their positions 
progressively: only a fraction $g$ of the cash is
invested in stock between $t$ and $t+1$ if the signal is to buy, and the same fraction $g$ of stock is sold if the signal is to sell. Typical values used below are $g \sim 1 \%$.

The normalized total order imbalance, which will be used to determine the change of price, is denoted
$Q(t)$:
\be
Q(t)= \frac{1}{\Phi} \sum_{i=1}^N q_i(t) = Q^+(t)-Q^-(t),
\ee 
where $\Phi$ is the total number of outstanding shares (that we assume to be constant), 
$Q^+$ is the fractional volume of buy orders and $Q^-$ the fractional volume of sell orders.

Agents sometimes choose to abandon their `chartist' strategies when the price
reaches levels that they feel unreasonable: when the price is too high, they
are likely to sell, and vice versa. More precisely, we construct a long term average of past
returns as:
\be
\overline{r}(t) = \frac{1}{1-\alpha} \sum_{t'< t} \alpha^{t-t'-1} \, r(t') 
\qquad 
r(t')=\log \left(\frac{X(t'+1)}{X(t')}\right),
\label{trace}
\ee  
where $\alpha < 1$ defines the time scale $T_0=1/\log(1/\alpha)$ over which the averaging is done, and $r(t)$ 
is the instantaneous stock return. When $\overline r$ is larger than a certain reference
return $\rho_0$, related to economy fundamentals, the stock can be deemed as overvalued
and `fundamentalists' will sell ($\epsilon_i=-1$). Conversely, if $\overline r < \rho_0$, the stock is
possibly undervalued, and $\epsilon_i=+1$. We model the occurrence of fundamental trading as stochastic, by
assigning a certain probability $p_f$ for every agent to follow a fundamental 
strategy rather than a technical (chartist) strategy. We want $p_f$ to increase with 
$|\overline r - \rho_0|$, and have chosen the following simple relation:
\be
p_f = \min\left(1,f \frac{|\overline r - \rho_0|}{\rho_0}\right), 
\ee
where $f$ is a certain parameter describing the confidence of agents in fundamental
information. Since $p_f$ increases when the price goes up too fast, fundamentalists have
a stabilizing role and give to the price a mean reverting component. In the following, we will assume that on the long run, the overall economy
growth $\rho_0$ and the interest rate $\rho$ are equal, and impose $\rho_0=\rho$, 
although in practice the two fluctuate with respect to each other. 

In the above rule, we have again assumed that all agents use the same time scale to determine 
the past average trend. This is probably very far from reality, where one expects that this time scale could be very different for different agents. 

Finally, it can be useful to consider the influence of `irrational' traders, who take
their decisions on the basis of random coin tossing only. We define $p_i$ as the probability 
for an agent to take a random decision. In this case, the probability to be a fundamentalist
is $(1-p_i)p_f$.
\vskip 0.3cm
\item{Price formation and market clearing mechanism: }
\vskip 0.3cm
Once the aggregate order imbalance $Q(t)$ is known, we update the price following
a simple linear rule \cite{BG,SantaFe,BC,CB,Farmer}: 
\be
r(t) = \log \left(\frac{X(t+1)}{X(t)}\right) \simeq\frac{X(t+1)}{X(t)} - 1= \frac{Q(t)}{\lambda},\label{price}
\ee
where $\lambda$ is a measure of the `stiffness' of the market. There has been recent 
empirical studies of this relation, which was shown to hold for individual 
stocks for small enough $Q$, on a sufficiently large time interval \cite{Gabaix}. For larger order imbalance, 
the price 
response appears to bend downward, a possible consequence of the structure of the order books. We have included this effect, with no noticeable effect on the qualitative results
presented below.

From a more microscopic point of view, i.e. on time scales smaller than the time unit that
we have chosen, agents place orders of different types in the market: market orders and limit orders. Market orders allows the order to be executed with certainty, but at the current 
market price. Limit orders ensures a maximum price for buy orders 
(and a minimum price for sell orders) but can be unexecuted, or 
only partially executed, depending on the
history of the price. Therefore, in general, the order put down by an agent will be only 
partially filled. We assume that the fraction of unfulfilled orders is the same for all
agents.
Market clearing is then ensured by the following rule: the global amount of sell orders
is $Q^-(t)$, and the total number of shares that can be bought at price $X(t+1)$
is:
\be
\tilde Q^+(t)=Q^+(t) \frac{X(t)}{X(t+1)}.
\ee
The fraction of filled buy orders $\varphi_+$ (resp. filled sell orders $\varphi_-$) is therefore:
\be
\varphi_+=\min\left(1,\frac{Q^-}{\tilde Q^+}\right) \qquad \varphi_-=\min\left(1,\frac{\tilde Q^+}{Q^-}\right).
\ee
From these quantities, one determines the actual number of shares $\delta \phi_i$ bought or sold 
by agent $i$:
\bea
&\delta \phi_i(t)& = g \varphi_+ \frac{B_i(t)}{X(t+1)} \qquad {\rm for} \ \epsilon_i(t)>0 \\ 
&\delta \phi_i(t)& = - g \varphi_- \phi_i(t) \qquad {\rm for} \ \epsilon_i(t)<0 
\eea
\vskip 0.3cm
\item{Wealth dynamics: }
\vskip 0.3cm
We now have all the ingredients to update the number of stocks and bonds 
of agent $i$, i.e.:
\bea \label{update}
\phi_i(t+1)&=&\phi_i(t) + \delta \phi_i(t)\nonumber\\ 
B_i(t+1)&=&B_i(t)(1+\rho)-\delta \phi_i(t) \, X(t+1) ,
\label{wealths}
\eea
where the last line adds the interest gained on the bonds between $t$ and $t+1$ to
the cash need to finance new stocks, or gained through stock selling. Note that
there is an {\it injection} of wealth due to the positive interest rate $\rho$. 
We will discuss this further in the following.
\vskip 0.3cm
\item{Update of the scores: }
\vskip 0.3cm
Each agent assigns scores to his strategies to measure their performance 
and uses at time $t$ the best strategy, i.e. the one with highest scores.   
We need now to specify  how the scores of the different strategies are updated.
The score of the $\alpha$th strategy of agent $i$ at time $t$ is denoted ${\mathcal S}_i^\alpha(t)$,
whereas the decision associated to this strategy when the available information at time $t$ is
${\mathcal I}_t$ is $\epsilon_i^\alpha({\mathcal I}_t)$. When the agent $i$ decides at time $t$ to
trade, the price at which the trade takes place is $X(t+1)$. Therefore, the {\it virtual}
profit he makes due to this trade is only known at time $t+2$ and is 
$\epsilon_i^\alpha({\mathcal I}_t) [X(t+2)-X(t+1)]$. 
We choose to update the score of the active strategies proportionally to 
the relative profit, corrected by the interest rate \footnote{A similar rule for the update of
scores was recently considered in \cite{Andersen-Sornette}}:
\be
{\mathcal S}_i^\alpha(t+1)=(1-\beta){\mathcal S}_i^\alpha(t)+
\beta \epsilon_i^\alpha({\mathcal I}_{t-1}) [r(t)-\rho], 
\quad \alpha=1,...,S-1,\label{scores}
\ee
whereas the score of the inactive strategy is identically zero. The parameter $\beta \leq 1$ 
defines a memory time: the performance of the strategies is only computed using the recent
part of the history. Note that the update of the scores is not weighted by the actual 
transaction volume: good decisions are valued independently of the current wealth of the
agent. Therefore, the score of the strategy is not proportional to the actual profit and
loss curve. 
 
One should keep in mind that only the best strategy $\alpha^*(t)$ is played by the agent 
at time $t$. Nevertheless, he updates the scores of all strategies {\it as if} they had
been played. In other words, market impact is neglected here, since the very fact of using
a given strategy influences the price itself. The history of the price would have been different
if a different strategy had been played. We do not take into account the market
impact for two reasons: first, the update of the score is delayed as compared to the action
itself (see Eq. (\ref{scores})) -- therefore, the main source of systematic bias 
discussed in the
context of the Minority Game in, e.g. \cite{Marsiliimpact}, is removed. Second, market impact
is in practice very hard to estimate for traders themselves (although some recent 
studies start addressing this issue \cite{Gabaix,Mantegna-Lillo,usunpub}),
and strategies are often backtested under the assumption that the market impact is small. 

A related point is that of the virtual profit computed above. Taking profit means closing
one's position, at a price that is not known in advance. Again there will be some market
impact and the actual price of the transaction is on average less than the current price. 
This effect is well known to active market participants and, as mentioned above,
has been recently the subject of some studies.
A way to model this is to add a transaction cost to the above update of the score,
independent of whether one buys or sells. (This cost should also
be taken into account in the above wealth balance).

\end{itemize}

\subsection{Summary of the parameters and main results}

The model contains a rather large number of parameters: the interest rate $\rho$,
the memory length used for technical trading $m$, the `polarization' of 
strategies $P$, the fraction of invested wealth $g$,
the time scale used to trigger fundamental trading $\alpha$, the propensity of 
fundamental trading $f$ and the fraction of irrational agents $p_i$, the stiffness 
of the market $\lambda$, and the memory time
of agents when the score of the strategies are updated $\beta$. However, the only truly important
parameters are $g/\lambda$ and the polarization $P$, which determine the 
qualitative behaviour of price changes. The other parameters
influence the quantitative results, but not the qualitative features, which is the
appearance of three qualitatively different regimes (see Fig. 1):

\begin{figure}
\begin{center}
\psfig{file=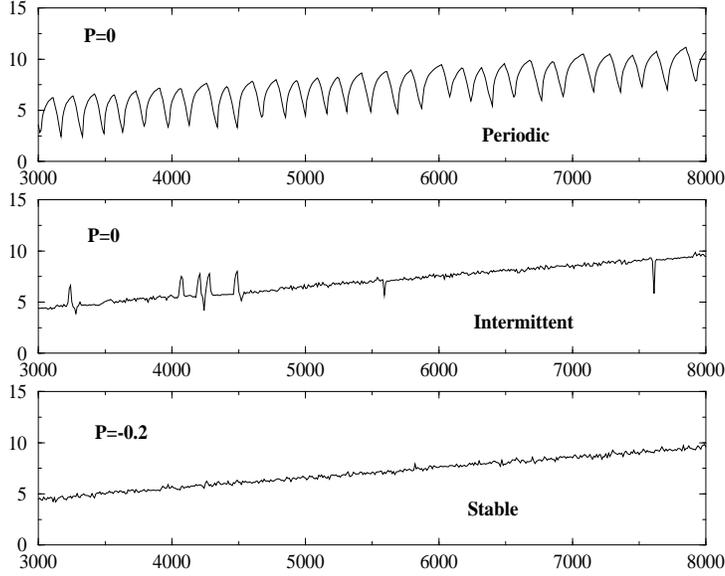,width=9.cm,height=6.5cm,angle=270} 
\end{center}
\caption{Typical price charts in the three regimes: periodic, intermittent, stable 
(efficient). 
The parameters are: $S=3$, $m=5$, $g=0.005$, $f=0.05$, $N=1001$,
$1-\beta=10^{-2}$ and $1-\alpha=10^{-4}$.
The top graph corresponds to $g/\lambda=0.1$, the two bottom graphs to
$g/\lambda=0.6$.}
\label{Fig1}
\end{figure}

\begin{itemize}

\item An {\it Oscillatory Regime}, corresponding to `weak coupling': 
$g/\lambda \siml 0.4$, and $P \geq 0$, where
speculative bubbles are formed, and finally collapse in sudden crashes induced
by the fundamentalist behaviour. In this regime, markets are not efficient, 
and a large fraction of the orders is (on average) unfulfilled.

\item A {\it Turbulent regime} ($g/\lambda \simg 0.4$, $P \geq -|P_0|$) where the `stylized' 
facts of liquid markets are well reproduced: the market is efficient (although
some persistent or antipersistent correlations survive),
the returns follow a power law distribution, and volatility clustering is present. 

\item A {\it Stable regime}, which arises if the polarization $P$ is sufficiently negative
(predominance of contrarian strategies). In this case, the fluctuations of the price are
mild and mean reverting (see Fig. 1 c), as one
would expect in a `rational market' where the trading price is always close to the 
fundamental price. 

\end{itemize}

\begin{figure}
\begin{center}
\psfig{file=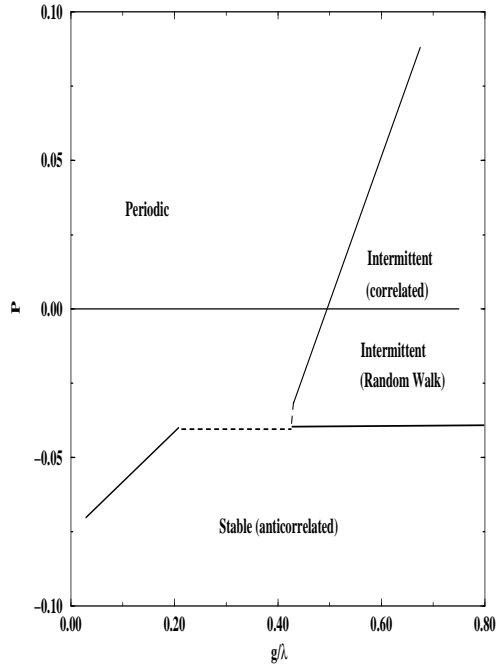,width=9.cm,height=6.5cm,angle=270} 
\end{center}
\caption{Phase diagram of the model. The region $g/\lambda > 0.4$,
$P \ll 1$ corresponds to 
an intermittent regime, with small linear correlations but strong volatility
fluctuations. The dashed lines correspond to crossover regions, where a mixed behaviour is
observed.}
\label{diagram}
\end{figure}

A (rather schematic) phase diagram of the model in the plane $(g/\lambda,
P)$ for a fixed value of all the other parameters is shown in Fig. \ref{diagram}.
This qualitative phase diagram is the central result of our study.

\section{Kinematics of the model: fully random strategies}

\begin{figure}
\begin{center}
\psfig{file=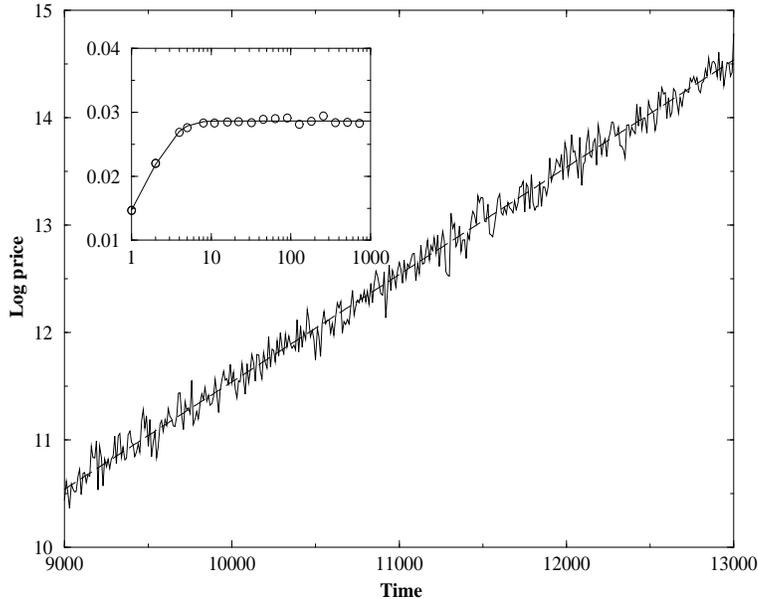,width=9.cm,height=6.5cm,angle=270} 
\end{center}
\caption{Behaviour of the price as a function of time for purely random strategies. 
Inset: Variogram of the price fluctuations, and Ornstein-Uhlenbeck fit.}
\label{Fig3}
\end{figure}

Before embarking to analyze the influence of strategies, it is important 
to calibrate the bare version of our model where agents take purely random,
uncorrelated decisions at each instant of time. In this case, the price
fluctuations will only reflect the wealth constraints. We show in Fig. \ref{Fig3} 
the price chart for some values of the parameters. The log-price performs a
mean reverting random walk around the fundamental price $X_f(t)=\exp (\rho t)$.
In the inset we show the log-price variogram, defined as:
\be\label{pricevariog}
{\mathcal V}(\tau)=\left \langle \left(\log \frac{X(t+\tau)}{X(t)} -\rho \tau\right)^2 \right \rangle,
\ee
together with an Ornstein-Uhlenbeck fit:
\be
{\mathcal V}(\tau)={\mathcal V}_\infty \left(1-\exp(-\tau/\tau_0)\right),
\ee
that describes a mean reverting random walk with a reverting time $\tau_0$,
and mean square excursion from the mean equal to ${\mathcal V}_\infty$. For small $\tau$, the 
behaviour of ${\mathcal V}(\tau)$ is linear in $\tau$, as for a free random walk, 
indicating that random trading leads, as expected, to unpredictable price changes. 
However, on larger time scales, 
the limitation of wealth and of stocks (agents cannot borrow nor short sell stocks) 
prevents the price from wandering infinitely far from the 
fundamental price, and leads to a mean-reverting behaviour. This mechanism will also operate 
for more complicated trading rules and will be discussed again in section 5.

It is simple to understand how these quantities depend on the parameters $g,\lambda$. 
From the price fixing mechanism, one can write a Langevin
equation for the price that reads:
\be
r(t)=\frac{d \log X}{dt} = \frac{g}{2 \lambda \phi} \left(\frac{B(t)}{X}-\phi +\xi(t)\right),\label{Langevin1}
\ee
where $B(t)$ is the average of an agent wealth in bonds, $\phi$ the average number of
stocks per agent, and $\xi$ is a random binomial variable measuring the small offset from a perfect 50-50 division 
between buyers and sellers. The variance of $\xi(t)$ is therefore of order
$N^{-1}$. Equation (\ref{Langevin1}) indeed describes a mean 
reverting process around $B(t)$: when the price is too high, the demand 
goes down due to budget constraints and the price goes down, and vice-versa.
The long term increase of price is only due, in our model, to the continuous injection of cash through the interest rate.  

Eq.~(\ref{Langevin1}) defines a mean reverting process, that allows one to obtain in particular:
\be
\tau_0 \propto \frac{\lambda \phi}{g} \qquad {\mathcal V}_\infty \propto 
\frac{g}{N\lambda \phi},
\ee 
in agreement with numerical results. The short time volatility of the
market, for $\tau \ll \tau_0$, is given by $\sigma^2 = {\mathcal V}_\infty/\tau_0
\propto g^2/N \lambda^2 \phi^2$, and is small for `stiffer' markets, or if the
fraction of invested wealth is smaller, as expected. 
Also, the volatility decreases when the number of agents increases. For a market 
with $10^4$ participants such that $g = 10\%$, this formula gives a 
reasonable volatility of $1 \%$ per week, when the market stiffness is
$\lambda \sim 0.1$. However, in this case, the return time is also of 
the order of a week and the total variability of prices is $1 \%$, 
both being far too small compared to reality.

Let us finally note that from the simulations, the market liquidity,
measured as the fraction $c$ of fulfilled orders, improves when the market 
stiffness $\lambda$ decreases. For example, we find $c=0.9$ for $g=1\%$, $\lambda=1$,
and $c=0.98$ for $g=1\%$, $\lambda=0.1$. 

\section{The oscillatory regime}

This regime is characterized by the presence of 
regular bubbles followed by rapid `crashes'. The period of the bubbles is a 
function of the model parameters. In Fig. 4 we show the dependence of the
period on ${g}/{\lambda}$, for some fixed values of the other 
parameters. The period  decreases as ${g}/{\lambda}$ increases, 
and vanishes when the market enters the turbulent regime. On the same
plot, we have also shown the fraction $c$ of fulfilled orders, which is 
very low in the periodic phase and increases with $g/\lambda$.

\begin{figure}
\begin{center}
\psfig{file=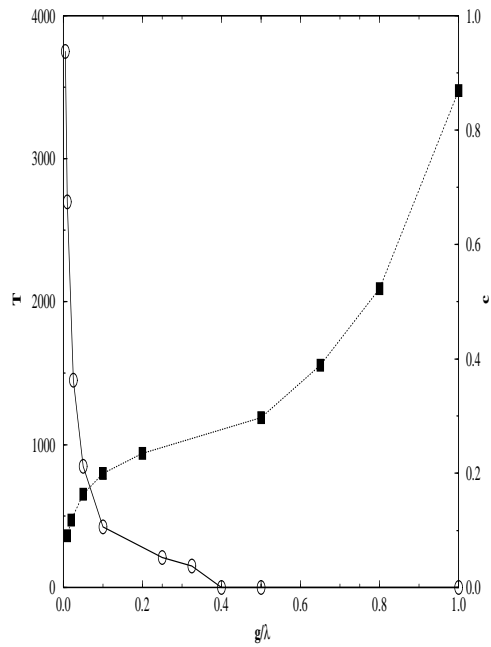,width=9cm,height=6.5cm,angle=270} 
\end{center}
\caption{Period $T$ of the oscillations (open circles, left scale), and
fraction $c$ of fulfilled orders (black squares, right scale), as 
a function of $g/\lambda$. All other parameters are as in Fig. \protect\ref{Fig1}.
}
\label{Fig4}
\end{figure}

\subsection{A qualitative discussion}

We want to understand the mechanism underlying the creation 
and persistence of bubbles, and their final collapse.

To answer this question, one has to 
look back at Eq. (\ref{price}) that describes the price dynamics. Since the 
strategies are randomly distributed among agents, we can expect that
at the beginning of the game one half of the agents is willing to buy and 
the other half to sell. The price increment is therefore
$r=\delta X/X \sim g\lambda^{-1} N/(2 \Phi) [B/X - \Phi]$. Since in general
$B/X \neq \Phi$, the price rises or decreases. Suppose that the initial
conditions are such that it increases. If $g/\lambda$ is small, both the
wealth and number of shares of the agents change quite slowly, and
therefore $B/X - \Phi$ will keep a constant sign for a while. This
will generate a history of price of the form $..,1,1,1,1,1....$ (See eq. 
(\ref{history})). This is the initial stage of the bubble. Now, the
population of agents fall into two categories: those who have at least one strategy 
such that $..,1,1,1,1,1$ sends a buy signal, and those who do not
have such a strategy available. The buy-strategies 
keep being rewarded, while the sell ones keep loosing points: agents who can 
buy will continue buying, while 
agents who cannot will soon become 
inactive since all active strategies have a negative score.  
In this way the bubble is self-sustained 
and the price keeps increasing. This `majority' mechanism was 
discussed in the context of the Minority Game in \cite{Marsili,Andersen-Sornette}. However, the buying power of the buyers 
decreases, because the price increases and the available cash decreases.
Hence, the relative return of the stock over the risk-free rate 
diminishes, and the score of the buying strategies only become marginally
positive (because the initial large gains are forgotten, only the 
recent past is included in the calculation of the scores). Since the 
return exceeds the reference rate $\rho_0$, a certain fraction of 
agents become fundamentalist and  act contrarily to the main trend, i.e. sell.
As soon as their action is such that the price drops, the history will
change to $..,1,1,1,1,-1$. At this point, one half of the active agents
still receive a buy signal, but the other half receive a sell signal 
(since strategies are random, and for now unbiased -- $P=0$). Now, 
these selling agents have a lot of stocks, since they have been buying 
for a long time, whereas the still buying agents have a poor buying power.
This obviously results in a series of $-1$, and an `anti-bubble' is created.
This anti-bubble has a large initial negative slope because the system
starts in a highly unbalanced state. There is no symmetry between a 
bubble and an antibubble because of the presence of a non zero interest
rate, which is a source of cash and favors, on average, positive trends. 
Once again, it is the presence of fundamentalist agents which at some point 
triggers the end of the descending trend, and re-establish a bubble.
The precise time at which the bubble collapses is random, and the way
the collapse is triggered is very similar to a nucleation process (see next
subsection).
 
One interesting quantity is the fraction of fulfilled orders. In the
oscillating regime, one understands from the above arguments that the unbalance between
 buy orders and sell orders is in general very large, leading 
to great amount of unfulfilled orders. The fraction $c$ of fulfilled
orders increases with $g/\lambda$: see Fig. 4. As will be discussed below, this is a driving 
force to escape from this oscillating regime, which obviously does not look at all
like real markets. 

The effect of the polarization $P$, which  can be appreciated in the 
phase diagram of Fig. \ref{diagram}, is quite easy to understand in the light of this 
microscopic analysis: a positive value
of $P$ increases the average number of trend followers and therefore is 
almost irrelevant for the bubble dynamics which 
is based on an endogenous trend following behaviour. On the other hand
$P<0$ acts contrarily to the bubble creation forcing a percentage of 
agents to act against the trend. Interestingly enough, as it can be seen from
 the phase diagram, even a small $P<0$ is able to prevent the appearance of
a bubble. As a function of $|P|$, the transition between an oscillating 
behaviour and a stable behaviour is first-order like, in the sense that the
period of the oscillation is still finite when the transition occurs. 

The above mechanism also allows us to understand the role of the parameters $f$ or 
the fraction of irrational agents $p_i$. The numerical simulations indicate that
increasing the value of these parameters may at first 
{\it stabilize} the oscillating phase, while one would intuitively expect an 
opposite  behaviour since fundamentalists/irrational players reduce the relative 
number of trend followers. This effect can be explained in the following way.
We have seen that, as the price increases during the bubble, 
the available cash of active players  and therefore their buying power 
decreases bringing the bubble to saturation. 
However, each active agent has a certain probability proportional to $f$ to
become a fundamentalist and sell, increasing his wealth in cash.
The buying pressure can then stay higher  for a longer 
period of time, determining more stable bubbles. 
Of course, as $f$ increases substantially, its other effect of nucleating 
opposite trends becomes the most important one, 
and the system behaves much as if the polarization parameter $P$ was
strongly negative, and thus prevents the appearance of bubbles. 
Finally, the presence of irrational agents who buy or sell randomly,
and therefore statistically drop out of the order imbalance, can be seen
as reducing the effective value of $g$, and thus -- somewhat surprisingly -- 
stabilizing the oscillating regime. 
We now turn to a mathematical transcription of the above discussion, that
allows us to characterize more accurately the {\it shape} of the bubbles
in our model.

Note finally that the oscillating regime tends to die out for very long times. This is
due to the fact that agents that have a buying strategy during the bubble tend to 
underperform on the long run. Their wealth is, at long times, insufficient to 
sustain the bubble. In order to determine the oscillation period more precisely, we 
have artificially given to each agent both its initial strategy and its perfect 
mirror image. This prevents the appearance of two groups with systematically different
wealths, without changing the basic mechanism leading to bubbles and crashes.

\subsection{A mean-field description}

In this section we  try to describe the bubble-crash dynamics by means of some 
mean-field equations. As we shall see, this will enable us to understand 
the precise mechanism of the bubble saturation and the consequent crash occurrence.

We start by  defining the relevant mean-field variables: the average amount $B$ 
of bonds and the average stock amount $\phi$ hold by active agents. 
It is also useful to define the number of buyers/sellers $N_b,N_s$.

Our aim is to write some evolution equations for these average quantities and for the price. 
In the following we shall consider the continuous time limit of equations (\ref{price}),
(\ref{wealths}) and (\ref{trace}). We have:
\be
r(t)=\frac{d\log X}{dt}= \frac{1}{\lambda} Q(t)
\ee
\be
\overline{r}(t)=\frac{1}{T_0}\int_0^t d\tau \exp(-\frac{t-\tau}{T_0})\, r(\tau),
\ee
where $T_0=1/\ln(1/\alpha)$ (see eq. (\ref{trace})). On the other hand the 
precise expression of the global action $Q(t)$ as well as  the wealths dynamics is not 
the same
during the whole bubble-crash cycle and to write down appropriate equations 
it is necessary to distinguish  different regimes where the agents behaviour can be 
considered 
as homogeneous in time.
In general  we can assume that during the bubble all the active players are buyers, 
except for a
small percentage of fundamentalists who  start selling when the price increases too much. 
On the 
contrary, during the crash all active agents become sellers while the fundamentalists will 
eventually act as buyers. In the following, we shall consider the case where the parameters
$g/\lambda$ and $f$ are very small. 

\subsubsection{Beginning of the bubble: $\overline{r}(t)<\rho$}

At the beginning of the bubble the price starts increasing but is still lower than 
the fundamental reference value $\exp(\rho t)$. This implies that all active agents,
including the fundamentalists, are buyers. The total number of buyers is 
therefore given by all the agents who have at least one buying strategy, plus a
stochastic contribution given by the fundamentalists:
\be
N_b=N(1-\frac{1}{2^s})+\frac{N}{2^s} p_{f},
\ee
where $s=S-1$. On the other hand, the active sellers rapidly disappear as the score of
the corresponding strategy deteriorates. The price rises without transaction, but since
$g/\lambda$ is small, it does so rather slowly. 
since all players who have not a buying strategy prefer
to remain inactive. In this context, it is evident that no transaction can be realized at all 
since there is a complete unbalance between offer and demand, and  the parameter $\varphi_{+}$, which 
determines  the fraction of fulfilled orders, is therefore identically zero.
The evolution equations are in this case very simple (see eq. (\ref{price})(\ref{wealths})):
\be
r(t)= \frac{g}{\lambda} N_b \frac{B}{\Phi X}
\ee
and:
\be 
\frac{d B}{dt}=\rho B \qquad \frac{d\phi}{dt}=0
\ee
The initial rate of growth of the bubble $r_0$ is therefore given by: 
\be
r_0 \approx \frac{g}{\lambda}\frac{B_0 N}{X_0 \Phi}(1-\frac{1}{2^s}),
\ee
where $B_0=B(t=t_0)$ and $X_0=X(t=t_0)$, and $t_0$ is the start date of the bubble. 
This gives a good approximation, as can be seen
from Fig. \ref{fit}. Since the value of $X_0$ just after the crash is small, the
buying power is high and $r_0$ is much larger than $\rho$. The value of $\overline{r}(t)$ 
therefore steadily increases.
 
\subsubsection{Saturation of the bubble: $\overline{r}(t)>\rho$}

\begin{figure}
\begin{center}
\psfig{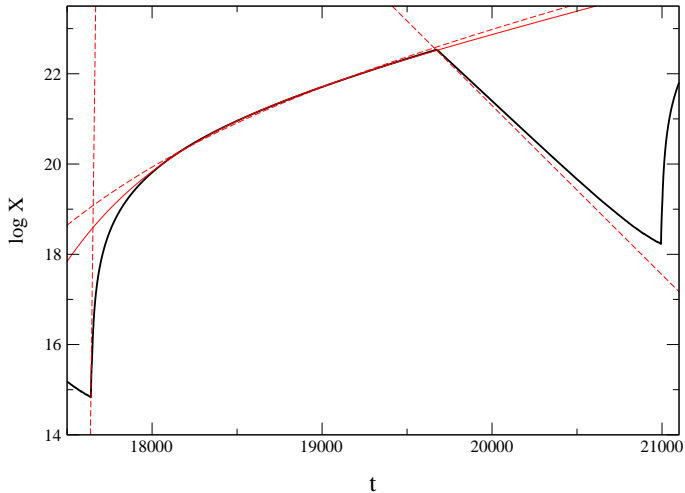} 
\end{center}
\caption{Bubble-crash dynamics in the oscillatory regime for $g/\lambda=0.005$, 
and $\rho=0.001,f=0.005, \alpha=0.9999$. The straight line corresponds to 
numerical data, the dashed lines to the mean-field approximations 
(see text), and the dashed-dotted line to the best exponential fit in the
bubble saturation regime.}
\label{fit}
\end{figure}

In this regime the fundamentalist players start acting contrarily to the main trend
and take their profit, thereby reestablishing some trading activity.  We now have:
\be 
N_b=N(1-\frac{1}{2^s})(1-p_f)  \qquad N_s=Np_f
\ee
and the return and wealths evolution equations now read
\be
r(t)= \frac{g}{\lambda \Phi}\left(N_b \frac{B}{X}-\frac{N_s}{N}\Phi\right)
\label{sat1}
\ee
\be
\frac{d B}{dt}=\rho B - (1-p_f) g \varphi_+ B + p_f g \phi X
\ee
\be
\frac{d \phi}{dt}=(1-p_f) g \varphi_+ \frac{B}{X}-p_f g \phi
\label{sat2}
\ee
where we have assumed that $\varphi_+ =  p_f \Phi X / N_b B \ll 1$, which is
consistent when $f \ll 1$. In this limit, the equations simplify significantly, since
all terms containing $p_f$ can be neglected. Introducing the buying power $\Upsilon=B/X$, one
has:
\be
r \approx \frac{gN}{\lambda \Phi} (1-\frac{1}{2^s}) \Upsilon \qquad 
\frac{d\Upsilon}{dt} \approx (\rho-r) \Upsilon,
\ee
from which one extracts: 
\be
\frac{dr}{dt}=(\rho-r)r \longrightarrow r(t)=\frac{\rho}{1-A \exp(-\rho t)},\label{sat22}
\ee
with $A=(r_0-\rho)/r_0 > 0$. This shows that $r(t)$ tends for long times 
to the overall growth rate of the 
economy. This is
expected since a faster growth cannot be sustained because of budget constraints:
the buying power would then tend to zero. We also find that $r(t)$ tends to 
$\rho$ exponentially, with a decay rate $\gamma$ equal to $\rho$ itself. 
A better approximation for $\gamma$ can be obtained by retaining the 
small terms dropped in the above analysis. In figure 
\ref{fit} we show the curve obtained from Eq.~(\ref{sat22}) with the value of $\gamma$ 
estimated analytically together with the best exponential fit of the data, and
find very good agreement.

\begin{figure}
\begin{center}
\psfig{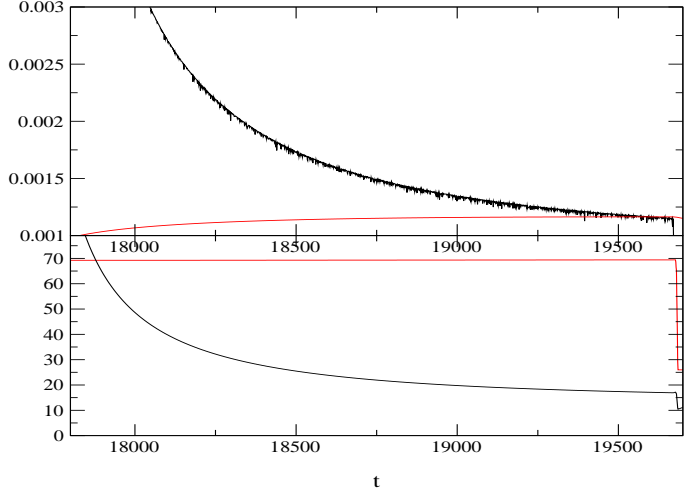} 
\end{center}
\caption{Upper graph: Behaviour of $r(t)$ (note the fluctuations) and $\overline{r}(t)$ 
as a function of $t$ in the bubble-saturation regime (respectively, 
straight and dashed lines). 
Lower graph: For the same time-interval, the buying power $B/X$ (straight line) and $\phi$ 
(dashed line). Same values of parameters as in  Fig. \ref{fit}. }
\label{sat}
\end{figure}

In this limit where $\rho T_0 \gg 1$, one can show that $\overline{r}(t)-\rho$ hardly 
varies during the bubble period, and is approximately equal to $A \rho$. 
We show in Fig. (\ref{sat}) the time evolution of various quantities during this bubble. 

The knowledge of $\gamma$ also allows us to estimate the time of occurrence of the 
crash. We know that the bubble is sustained because buying strategies keep being rewarded when
the price keeps increasing. Nevertheless, we have seen that during the bubble saturation 
the average return approaches the reference value $\rho$,
thus also the performance of buying strategies (i.e. their scores relative to 
the risk free rate) will increase less as time goes on; a small drop of price might 
then be enough to both trigger some strong selling strategies and favor the inactive
strategy, reducing the buying pressure. In the absence of fluctuations, the average role 
of the fundamentalists is to renormalize the values of $A$ and $\gamma$ in Eq.~(\ref{sat22})
above. The number of fundamentalists is however a stochastic Poisson process, and
fluctuates around the average value $Np_f$ with variance $\sigma_f = \sqrt{N p_f}$. 
One therefore expects that when $r(t)-\rho$ becomes of the order of $g \sigma_f/\lambda N$,
one of such positive fluctuations will bring the instantaneous return below $\rho$ 
and the price history will therefore change from $1,1,1,...,1$ to $1,1,1,...,-1$, from 
which a sudden wave of sell orders triggers the crash.
A reasonable estimate of the bubble end-time $t^\star$ is therefore given by:
\be
\rho A \exp\left(-\gamma (t^\star-t_0)\right) \approx \frac{g}{\lambda} \sqrt{\frac{p_f}{N}},
\ee 
or
\be
t^\star-t_0 \approx \frac{1}{2\gamma} \log \left(\frac{A \lambda^2 \rho^2 N \gamma T_0}{f g^2}\right)
\label{tempocrash}
\ee
which indeed has the correct order of magnitude and 
qualitative behaviour, increasing with decreasing  $g/\lambda$ (see Fig. 4). 
Note also that
as $f \to 0$, the bubbles have an infinite lifetime, as we have seen in our simulations. 
Some fundamentalist behaviour is needed to nucleate a crash. 

\subsubsection{Crash}

At the beginning of the crash all the fundamentalists act as 
sellers since one still has $\overline{r}(t)>\rho$. Thus
\be
N_b=0  \qquad N_s=N(1-\frac{1}{2^s})+N p_f\frac{1}{2^s}
\ee
and the initial slope of the crash is now given by:
\be
r^\star  \approx - \frac{g}{\lambda} \frac{N}{\Phi} (1-\frac{1}{2^s})
\label{crash}
\ee

When at some point $\overline{r}(t)$ becomes smaller than $\rho$  
the fundamentalists start acting again contrarily 
to the main trend driving back the price toward the fundamental value $X_f(t)$. 
We may expect a saturation  effect as in 
the previous regime, which would enable an asymptotic expansion. 
However, as it can be seen from 
the figures, the crash period is much shorter than the bubble one and the crash 
stops much before any asymptotic regime set in. The reason for this asymmetry is the
interest rate that `refills' the buying power in the bubble regime, that has no
counterpart in the crash regime. 
The crash ends in the same way as the bubble does, through a nucleation process
when $r(t) - \rho$ becomes of the order of the fluctuations (which are in this case 
much larger).

\section{The intermittent regime}

\subsection{Results from the simulation} 

\begin{figure}
\begin{center}
\psfig{file=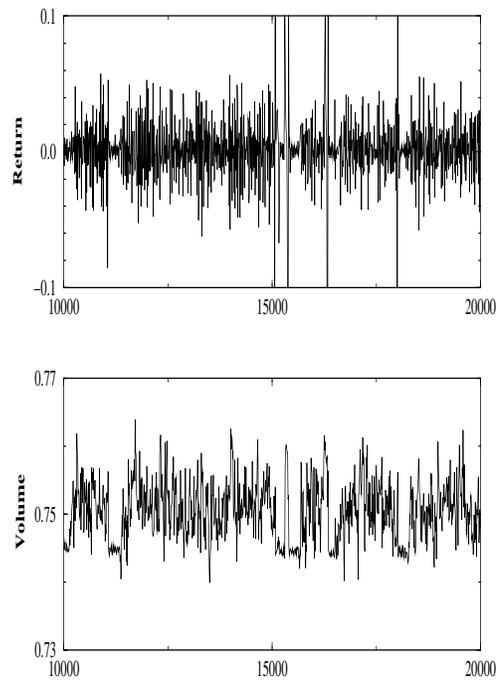,width=9.cm,height=6.5cm,angle=270} 
\end{center}
\caption{Time series of the returns, truncated to $\pm 10\%$ (top panel), 
and of the fraction of active agents (bottom panel), 
for $g/\lambda=0.75$.}
\label{FigI1}
\end{figure}

Upon increasing the parameter $g/\lambda$, the oscillating regime disappears and gives rise to 
an interesting market behaviour, where different market `states' coexist: bubbles and 
crashes, periods of very small activity, and periods of very large activity, intermixed with
each another. A typical chart of the returns and of the volume as a function of time is plotted
in Figure \ref{FigI1}. From visual inspection, it is quite clear that the return time 
series exhibits volatility clustering.
We now turn to a more quantitative analysis of the price series statistics. 
The data we analyze below corresponds to $N=10000$ agents, 
with `unpolarized' strategies $P=0$, and for $g/\lambda=0.75$, 
deep in the intermittent regime (see Fig. \ref{diagram}). 
Other parameters are identical to those in Fig \ref{Fig1}. We note that the qualitative effects we report 
below have not been seen to depend on $N$, at least up to $N = 10000$ which is the
largest size we have investigated. Strong size effects, reported in the Lux-Marchesi model 
\cite{Stauffer} for instance, seem to be absent in our case. However, we expect that 
when $N$ becomes comparable to the total number of strategies (i.e. $2^{2^m}$), 
the phenomenology will change since many agents will share exactly the smae strategies.

\begin{figure}
\begin{center}
\psfig{file=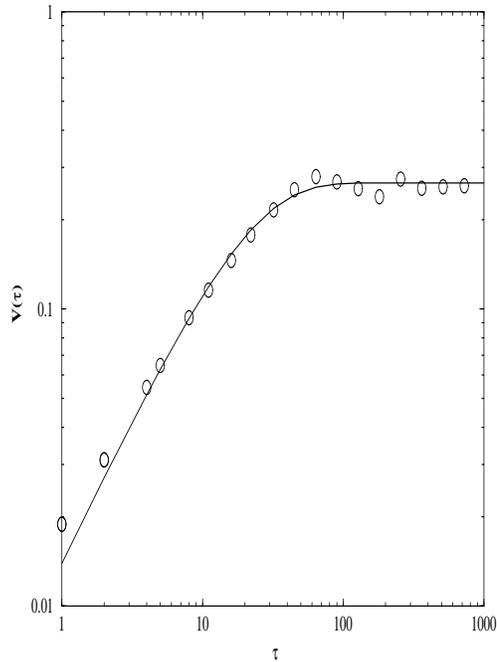,width=9.cm,height=6.5cm,angle=270} 
\end{center}
\caption{Variogram of the price as a function of the time lag, together with an Ornstein-Uhlenbeck
(mean-reverting) fit.}
\label{FigI2}
\end{figure}

First, we look at the price variogram, defined by Eq.(\ref{pricevariog}). 
This is shown in Fig.
\ref{FigI2}, together with an Ornstein-Uhlenbeck fit. The saturation time $\tau_0$ is
of the order of 100; taking the unit time in our model to be the week, this corresponds to 
one or two years, with a volatility of roughly $20 \%$ per year, which is quite 
reasonable. As explained in section 3 the saturation is a direct consequence of the bounded wealth of agents. Is there a similar mean-reverting trend in real markets ? The analysis of the variogram of the Dow-Jones index, for example, in the 
period 1950-2000, shows no convincing sign of saturation on the scale of the year, although a slight
bend down wards for longer time lag is visible, but the data become noisy. There has been reports 
in the literature of a systematic mean-reverting effects on the scale of 5-10 years, perhaps related to 
the mechanism discussed here. Remember that the fundamental price in our model has zero volatility.
The short time volatility is the result of pure trading, which leads to a random walk like behaviour of the price; the saturation occurs because of insufficient 
resources to sustain a large difference between the fundamental price and the speculative price.

\begin{figure}
\begin{center}
\psfig{file=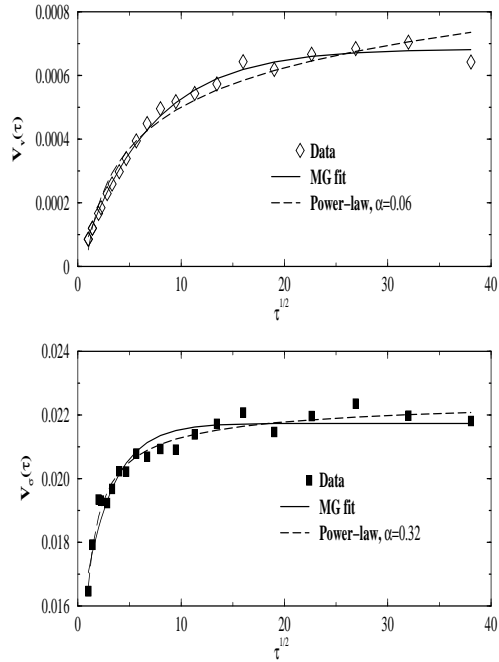,width=9.cm,height=6.5cm,angle=270} 
\end{center}
\caption{Variogram of the volume (top panel) and of the absolute return (Bottom panel) as a function of the time lag, together with the fits given by Eq.~(\protect\ref{variogfit}) and  Eq.~(\protect\ref{variogfitpwr}).}
\label{FigI3}
\end{figure}

The two other quantities that we have systematically studied are the volume variogram and the 
absolute return variogram, defined as:
\be\label{variog}
{\mathcal V}_{v,\sigma}(\tau)=\left \langle \left(O(t+\tau)-O(t)\right)^2 \right \rangle,
\ee
where $O(t)$ denotes, respectively, the fraction of active agents and the absolute return $|r(t)|$.
We show these two quantities in Fig. \ref{FigI3} as a function of $\sqrt{\tau}$, together with a fit inspired from the theory explained in the next subsection:
\be\label{variogfit}
\left.{\mathcal V}(\tau)\right|_{SQRT}= {\mathcal V}_\infty \left(1 -\exp(-\sqrt{\frac{\tau}{\tau_0}})\right).
\ee 
Note that the short time behaviour of the above function is $\sqrt{\tau}$, in contrast with the
regular (linear) behaviour of the Ornstein-Uhlenbeck form. We will explain the origin of this
singularity below, and explain why Eq.~(\ref{variogfit}) fits very well the volume variogram. 
For the variogram of the absolute returns, we have added to Eq.~(\ref{variogfit}) a non zero 
constant, that takes into account the fact that $|r(t)|$ is a noisy estimate of the volatility;
this extra noise is uncorrelated for different days and adds a contribution 
proportional to $1-\delta_{\tau,0}$. We have also shown a power-law fit
\be\label{variogfitpwr}
\left.{\mathcal V}(\tau)\right|_{PWR}= {\mathcal V}_\infty \left(1 -\left(\frac{\tau}{\tau_0}\right)^{-\alpha}\right),
\ee
which has been advocated in many empirical studies, with $\alpha \sim 0.1 - 0.3$. 
As can be seen from
Fig. \ref{FigI3}, the two fits are of comparable quality. 
Note that the value of $\alpha$ found for ${\mathcal V}_{v}(\tau)$ 
is significantly smaller that that for 
${\mathcal V}_{\sigma}(\tau)$, as also found for real market data. The reason for this will be
explained in the next subsection.

\begin{figure}
\begin{center}
\psfig{file=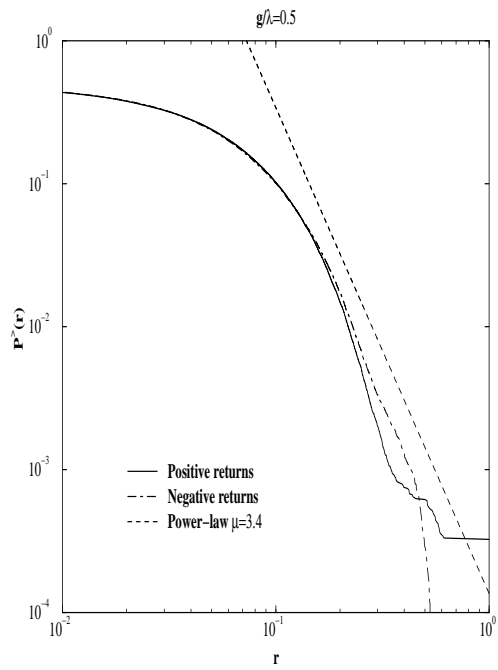,width=9.cm,height=6.5cm,angle=270} 
\end{center}
\caption{Distribution of returns for $g/\lambda=0.5$. A power-law tail with $\mu=3.4$ is shown for
comparison.}
\label{FigI4}
\end{figure}

We have also studied the distribution of returns. 
Not surprisingly, this distribution is found to be highly kurtic, which is expected since the volatility is fluctuating. The tail of the distribution 
can be fitted by a power-law, with an exponent $\mu \approx 3.5$, similar to 
the value reported in 
\cite{Stanley}: see Fig \ref{FigI4}. 
Note that the negative tail is slightly fatter than the positive tail.

The role of a non zero polarization of strategies $P$ is, for small enough $P$, to
induce some correlations (or anticorrelations) in the returns. For $P<0$, as was the
case for the oscillating regime, there is a first-order (discontinuous) transition 
toward a stable market, with small, strongly anticorrelated fluctuations that
track the fundamental price. This transition occurs for rather small values of 
$|P|\approx 0.03$. For $P > 0$, on the other hand, the intermittent phase
survives but the variogram of price fluctuations shows significant positive correlations.
For sufficiently large $P$, the oscillating phase reappears in a continuous way.

The conclusion of this subsection is that volatility clustering appears for large values of $g/\lambda$.
Qualitatively similar effects are seen for different choices of $m$ (memory time for the strategies), $\beta$ (memory time for the scores), and $S$ (number of strategies per agent), provided $g/\lambda$
is large enough to be in the intermittent phase. A crucial ingredient, however, is the existence 
of an inactive strategy, i.e., the fact that the volume of activity is allowed to fluctuate. We have
not been able to obtain long term volatility correlations of the type reported in Fig. \ref{FigI3} 
when all strategies are active. This observation has motivated us to propose a simple mechanism for non trivial volume (and volatility) fluctuations 
\cite{QF,PhysicaA}, that we discuss now in the present context.

\subsection{A simple mechanism for long-ranged volume correlations}

\subsubsection{Random time strategy shifts}

In the above model, as in the Minority Game, scores are attributed by agents to their 
possible strategies, as a function of their past performance. In particular, the inactive
strategy is adopted when the score of all active strategies are negative. 

In the turbulent regime we have seen that the market is 
`quasi-efficient': the autocorrelation of the price increments is 
close to zero. To a first  approximation no strategy can on the long 
run be profitable. This implies that the strategy scores locally behave, as a function of  time, as random walks. This very fact enables us to explain the fluctuations
of volume. Let us consider for simplicity the case $S=2$ (one active strategy and one 
inactive strategy per agent).
Since the switch between two strategies occurs when 
their scores cross, the activity of an agent is determined by the survival time of the 
active strategy over the inactive one, that is by the return time of a  
random walk (the score of the active strategy) to zero.  The interesting point is that 
these return times are well known to be power-law distributed. This leads 
immediately  to the non trivial behaviour of the activity variogram 
shown in Fig.\ref{FigI3}. The same argument was used to explain the volume fluctuations
in the Minority Game with an inactive strategy, in the efficient phase \cite{QF}.

More formally, let us define the quantity $\theta_i(t)$ that is equal to $1$ 
if agent $i$
is active at time $t$, and $0$ if inactive. The total activity is given by 
$v(t)=\sum_i \theta_i(t)$, and the activity variogram is given by 
\be
V_v(t,t')=\langle \left[v(t)-v(t')\right]^2 \rangle= 
C_v(t,t)+C_v(t',t')-2C_v(t,t').
\ee
where $C_v(t',t)$ denotes the volume correlation function.

One can consider two extreme cases which lead to the same result, up to a 
multiplicative constant: (a) agents follow 
completely different strategies and have independent activity patterns, i.e.
$\langle \theta_i\theta_j\rangle \propto \delta_{i,j}$ or (b) agents follow 
very 
similar strategies,  in which case $\theta_i=\theta_j$. 
In both cases, $C_v(t,t')$ is proportional to $\langle \theta_i(t)\theta_i(t') \rangle$ and can be exactly expressed in terms of the distribution $P(s)$ of the survival 
times $s$ of the active strategies \cite{gl,QF}. For
an unconfined random walk, the return time distribution $P(s)$ decays 
as $s^{-3/2}$ for large $s$. Note that this $s^{-3/2}$ behaviour is super-universal and only
requires short range correlation in the score increments, not even a finite 
second moment \cite{Sparre,Frisch}. However, in our model, the finite memory 
with which the scores are updated (i.e. the value of $\beta < 1$ in Eq.~(\ref{scores}))
leads to a truncation of the $s^{-3/2}$ beyond a time $\tau_0 \simeq 1/\log(1/\beta)$.
Without this truncation, the volume $v(t)$ would never become a stationary process, i.e.
$C_v(t,t')$ would still depend on both $t$ and $t'$ at long times, a phenomenon called 
`aging' in the physics literature \cite{aging}. For finite $\tau_0$, one can show analytically that  \cite{QF,PhysicaA}:   
\be
V_v(t+\tau,t) = V_v(\tau) \propto  \sqrt{\frac{\tau}{\tau_0}} \qquad \tau \ll \tau_0.
\label{cluster}
\ee
The formula given by Eq.~(\ref{variogfit}), was found to reproduce very well the 
crossover from this exact short time singular behaviour to the saturation regime.

The interesting conclusion is the following: the very fact that agents compare the performance of 
two strategies on a random signal leads to a multi-time scale behaviour of the
volume fluctuations. This argument accounts very accurately for our numerical data both for
the Minority Game and the present market model (see Fig. \ref{FigI3}) \cite{QF,PhysicaA}, and also reproduces quite well the empirical variogram of activity 
in real markets \cite{QF}.

\subsubsection{Volume and volatility}

Let us now discuss the relation between volume and volatility. In real markets, the two are known to be correlated; more precisely, it has recently been shown in \cite{Gopi} that the
long run correlations in volatility come from the long range correlation in the 
volume of activity (see also \cite{Bonnano}). In our artificial market, a scatter plot of 
the logarithm of the absolute return, $\log |r(t)|$ versus the volume of activity 
$v(t)$ shows nearly 
no correlations when the volume has small fluctuations around its average value $\overline{v}$, 
but is strongly correlated with $v(t)$ when $v(t)$ has large excursions above $\overline{v}$. This means that periods of high activity are also periods of large
volatilities. One therefore expects that the structure of temporal correlations of
the volume discussed above is also reflected in the volatility. If the relation 
between $|r(t)|$ and $v(t)$ was linear, or weakly non linear, one would in fact 
expect exactly the same shape for the variogram. The fact that this relation 
is highly non linear, i.e., nearly no correlations for $v(t)\approx \overline{v}$,
and a roughly exponential relation for $v(t) >  \overline{v}$, adds an instantaneous noise contribution proportional to $1-\delta_{\tau,0}$ to the variogram of absolute returns, and leads to a strong distortion of the shape of the relaxation. The fact that the effective power-law $\alpha$, defined in Eq.~(\ref{variogfitpwr}), is larger for $|r(t)|$ than
for $v(t)$ can be understood in details in the context of the multifractal random walk model 
of Bacry et al. \cite{MRW}.

\subsubsection{Other mechanisms for long-ranged correlations}

Recently, many agent based models have been proposed to account for the
stylized facts of financial markets, in particular volatility clustering and 
long range dependence \cite{Huberman,Lux1,Lux2,Kirman,Raberto,Bornholdt,Alfarano}. 
From the analysis of these models, one can distinguish 
three main mechanisms for this long-range dependence:

\begin{itemize}

\item Subordination of the strategies to performance. This is the mechanism 
explained above: as soon as each agent has different strategies with different
levels of activities, and that the choice between these strategies is subordinated to
their performance, one expects to see long range dependence of the type described above,
whenever these strategies lead to identical long term performance.
A similar mechanism is found in the models of \cite{Huberman,Lux2,Kirman}, where agents 
switch between different trading styles (e.g. fundamentalists/chartists) as a function 
of their perceived performance and of herding effects. The basic prediction of this 
scenario is the short time square root singularity of the volume variogram. This
prediction is very well obeyed in the Lux-Marchesi model, as shown in Fig. \ref{FigLux}.

\begin{figure}
\begin{center}
\psfig{file=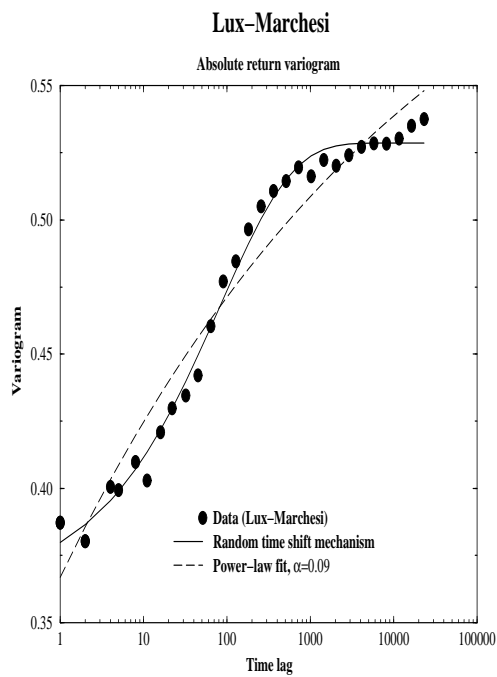,width=9.cm,height=6.5cm,angle=270} 
\end{center}
\caption{Variogram of the absolute returns as a function of the time lag in the
Lux-Marchesi model with parameters as in \cite{Lux2}. We compare with the fit 
suggested by the random time strategy shift mechanism (Eq.~(\protect\ref{variogfit})) and
a power-law fit Eq.~(\protect\ref{variogfitpwr}), as suggested by multifractal models \cite{MRW}.}
\label{FigLux}
\end{figure}

\item Subordination of the volatility to the price. In many models, the level of 
activity depends on the difference between the current price and a fundamental price.
For example, in the model considered by Bornholdt \cite{Bornholdt}, the volatility is a growing 
function of the absolute difference of these prices. In the model recently studied by 
Alfarano and Lux \cite{Alfarano}, on the contrary, the volatility is a decreasing function of this difference. 
In these models, the price is mean reverting toward the fundamental price. Calling $y(t)
=\log (X(t)/X_f(t))$, where $X_f$ is the fundamental price, a schematic equation for $y(t)$ is:
\be
\frac{dy}{dt}= - \kappa y + \sigma(y) \xi(t),
\ee
where $\xi$ is a white noise and $\sigma(y)$ a certain function. 
The corresponding temporal correlation function of the volatility can, for some 
specific forms of $\sigma(y)$, be exactly computed, and generically leads to a non 
exponential decay that can mimic long term dependence. Empirically, one can indeed
detect, on the Dow Jones index over a century, some correlation between the volatility 
and the difference between the current level and the average level of the index. However,
empirical studies show that volatility clustering exists even when the market is close 
to its average level. 

\item Heterogeneity of the agents time scales. Another mechanism, close in spirit to the 
`HARCH' model \cite{harch} or the cascade models proposed recently 
\cite{Ghash,Mandelbrot,Fisher,muzy,MRW}, comes from the different time
horizons used by the agents to set up their strategies. For example, in the model of 
Raberto et al. \cite{Raberto}, agents place orders at a distance from the current price proportional 
to a sliding average of the past volatility. The time scale used by the different agents 
is uniformly distributed between 10 day and 100 days. Correspondingly, this induces a
non exponential decay of the volatility correlation function when $\tau \le 100$. 

\end{itemize}

All these three mechanisms are expected to play a role in real markets. 
More precise statistical tests will hopefully allow one to distinguish between them 
and estimate their relative contribution to the long-range dependence effects. 

\subsection{Crashes: the role of memory and the dynamical freezing
of the choice mechanism}

We have also observed the following two interesting effects in the 
intermittent phase: 
\begin{itemize}

\item The frequency of bubbles/crashes decreases when the time horizon $m$ of the
strategies is increased.

\item  Just after a crash 
that follows a speculative bubble, the volatility is anomalously small.
\end{itemize} 

First, how do bubbles form in this intermittent regime ? The price fluctuations
are almost without correlations. In this context it happens 
with probability $1/2^{m+1}$  that at a certain time step the $m$-bit past 
price history is (1,1,....1) {\it and} simultaneously the buying 
power of buyers is larger than that of sellers. In this case, at the next time step the 
price again increases and the past history is mapped onto itself. Also, 
since $g/\lambda$ is large, the price change is substantial 
and the score of the buying strategies increases fast, thereby creating a
bubble that will terminate in the same way as we described in section 4. 
Thus, we expect an average frequency of $1/2^{m+1}$ for the bubble/crash occurrences, 
which one can easily verify in our model.

The second question concerns the period of very low volatility
which systematically follows the crashes.
In our simple model, we observe the presence of well
 defined cycles in the post-crash dynamics, which may vary for different 
numerical runs,  but are identical for all the crashes in the same 
simulation. The presence of these cycles obviously limits the amplitude of 
price fluctuations and reduce the volatility. Cycles occur because during the 
crash, the scores of selling strategies suddenly accumulate a large excess and 
are thus used for some time after the crash irrespective of their more recent 
performance. Thus all agents keep using the very same strategies. 
Since the number of possible past histories is finite ($2^m$), the dynamics 
is deterministic and the system must enter a cycle.

\section{Market efficiency and stability}

\subsection{Dynamical evolution of the parameters and self-organization}
\label{efficiency}
As discussed above, our model {\it a priori} involves a large number of parameters.
However, the precise value of most of them is not crucial. The only two important parameters
are the ratio $g/\lambda$, that measures the impact of trading on prices, and $P$, that
measures the tendency of the agents for trend following ($P>0$) versus contrarian strategies 
($P<0$). For small values of $g/\lambda$, the volatility of price changes is very small, 
and, as we have discussed in section 4, bubbles form in such a way that 
very visible trends appear. Furthermore, for small $g/\lambda$, these bubbles have a
very long lifetime. This means that the agents will naturally increase the fraction 
of the wealth they invest in this market, since the perceived risk is small. Hence $g$ 
will spontaneously increase. Simultaneously, small $g/\lambda$'s lead to rather small
execution rates: see Fig. \ref{Fig4}. Therefore, the microstructure of the market will
evolve such as to make $\lambda$ smaller, so that the market becomes more liquid. 
Both effects lead to an increase of the ratio $g/\lambda$, and the corresponding destruction
of clear trends. This scenario therefore suggests that $g/\lambda$ enters the intermittent
region, where the market becomes `quasi-efficient' (i.e. returns are uncorrelated on short time scales), but where interesting statistical anomalies appear. The ratio $g/\lambda$
then stops growing, since the market becomes very volatile, without clear trends. The agents 
therefore limit their investment. The above scenario is somewhat related to that proposed
in the context of the Minority Game in \cite{MG3}: increasing the number of players leads to
a more efficient game, but reduces the incentive for new players to enter, such as 
a `marginally efficient' state is reached. 

The role of the parameter $P$ is quite interesting. If agents were on average only 
slightly contrarian, the behaviour of the market would be completely different, with 
boring small mean reverting fluctuations around the fundamental price, that we have here
modeled as a deterministic growth $X_f(t)=\exp(\rho t)$. If this fundamental price was 
itself randomly fluctuating, our model market would be very close to the standard 
`efficient market' picture, where rational agents systematically correct past excess 
returns such as to lock the market price to the fundamental price. The fact that 
human psychology seems to favor mimetic, trend following strategies seems to keep the market 
in the intermittent region of the phase diagram shown in Fig. \ref{diagram}.

\subsection{Distribution of agents' wealth}

We have also studied the distribution of the wealth of the agents in our model. Interesting, for small values of $g$, the distribution is very uniform across agents: since the level
of speculative investment is small, the redistribution effect through trading dominates 
the random speculative gains. When $g$ increases, however, a Pareto like distribution of
wealth, with very unequal agents, sets in, as expected from the analysis of \cite{BM}.

\subsection{The effect of finite transaction costs}

An interesting question in the context of financial markets is to understand how the
introduction of transaction costs (as the Tobin tax for example) might stabilize the
behaviour of these markets. We have studied a generalization of our model in which 
a non zero proportional fee is paid at each transaction. As we shall see, this drives the system toward a much more stable regime. Transaction costs play a similar role to a
negative polarization (more contrarian) strategies.

The equations to be modified are the ones concerning the update of cash,
that accounts for the extra cost: Eq.~(\ref{update}) is changed into: 
\be
B_i(t+1)=B_i(t)(1+\rho)-\delta\phi_i(t) X(t+1)-\nu |\delta\phi_i(t) X(t+1)|.
\label{bond2}
\ee
Similarly, we adopt a modified update for the strategies scores (see \ref{scores}):
\be
S_i^{\alpha}(t+1)=(1-\beta)S_i^{\alpha}(t) + \beta \left(\epsilon_i^\alpha({\mathcal I}_{t-1})[r(t)-\rho] -\nu |\epsilon_i^\alpha({\mathcal I}_{t-1})|\right).
\label{scores2}
\ee

\begin{figure}
\begin{center}
\psfig{file=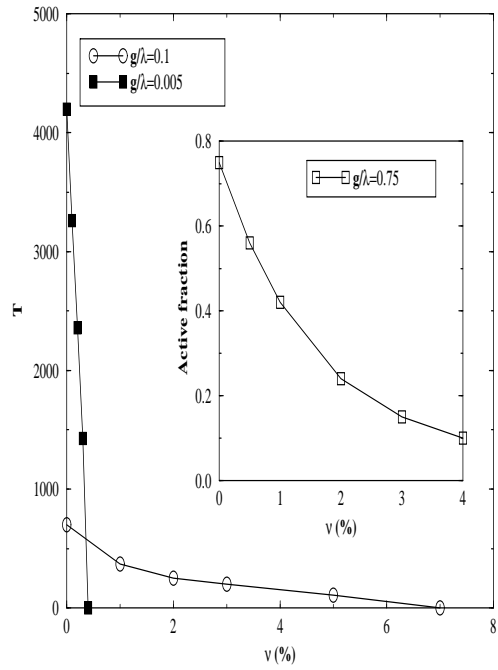,width=9.cm,height=6.5cm,angle=270} 
\end{center}
\caption{Evolution of the period $T$ of the oscillations as a function of the transaction
cost parameter $\nu$, for $g/\lambda=0.005$ and $g/\lambda=0.1$. Inset: Evolution of the
fraction of active agents as a function of $\nu$ for $g/\lambda=0.75$.}
\label{Figfee}
\end{figure}

Let us first discuss the small $g/\lambda$ regime, where bubbles appear in the
absence of transaction costs. Scores of the buying strategies grow most rapidly
at the start of the bubble, where the return $r(t_0)$ is maximum. When the transaction costs are such that $\nu > r(t_0)-\rho$, the bubble will never start. Since the initial growth of the bubble is proportional to $g/\lambda$, we expect that bubbles are more
robust to costs for larger $g/\lambda$'s. If a bubble is
formed, is  bound to collapse earlier for $\nu > 0$, since the performance of the buying 
strategy is systematically reduced. Hence the frequency of the oscillations will 
increase with $\nu$. We show in Fig. \ref{Figfee} that this is
indeed the case. (Note however that as $\nu$ increases, some bubbles do not form, and
one observes a mixed behaviour with on and off oscillations). In the intermittent phase, 
the introduction of transaction
costs favors the inactive strategy, and we observe a decay of the average fraction of 
active agents as $\nu$ increases (see Fig. \ref{Figfee}). Large values of 
transaction costs drive the system toward a more stable regime, 
that still however exhibits long-range volume fluctuations of the type described 
in the previous section.

The above discussion suggests that transaction costs would stabilize the markets. However, since the values of $g/\lambda$ that corresponds to realistic markets is rather large, 
these costs would have to be substantial (a few percent per trade) in order to really 
affect the markets. This conclusion could be read backward: a small tax, of the order
of a few basis points (i.e. $10^{-4}$), would probably not change dramatically the 
behaviour of markets, and would simply add to already existing costs (brokerage, 
bid-ask spreads and market impact).

\section{Summary and conclusion}

In this paper, we have presented a detailed study of a rather complex market model, 
inspired from the Santa Fe artificial market and the Minority Game. Agents have 
different strategies among which they can choose, according to their relative profitability,
with the possibility of {\it not} participating to the market. The price is updated 
according to the excess demand, and the wealth of the agents is properly accounted for. 
The set up of the model involves quite a large number of parameters. Fortunately, 
only two of them play a significant role: one describes the impact of trading on the 
price, and the other describes the propensity of agents to be trend following or contrarian. 

The main result of our study is the appearance of three different regimes, depending on the
value of these two parameters: an oscillating phase with bubbles and crashes, an intermittent 
phase and a stable `rational' market phase. The statistics of price changes in the intermittent phase resembles that of real price changes, with small linear correlations, fat tails and long range volatility clustering. We have discussed how the time dependence of these parameters could spontaneously lead the system in the intermittent region.
We have analyzed quantitatively the temporal correlation of activity in the intermittent
phase, and have shown that the `random time strategy shift' mechanism proposed in an earlier
paper \cite{QF} allows to understand these long ranged correlations. Other mechanisms 
leading to volatility clustering have been reviewed. 

We have discussed several interesting issues that our model allows to address, such as the
detailed mechanism for bubble formation and crashes, the influence of transaction costs, 
the distribution of agents wealth, 
and the role of a limited amount of capital on the long time fluctuations of the price. 

Many extensions of the model could be thought of, for example, the diversity of time
horizons used by the different agents, the introduction of time dependent fundamental
factors, or a larger number of tradable assets. A more interesting path, in our mind, 
would be to simplify the model sufficiently as
to be able to analytically predict the phase diagram of Fig. \ref{diagram}, and reach a
similar level of understanding as in the Minority Game (for some work in this direction,
see \cite{Marsili}.)

\section*{Acknowledgments} We thank M. M\'ezard, who participated to the early stages of
this work, for discussions. We have benefited from many conversations with S. Bornholdt,
A. Cavagna, D. Challet, D. Farmer, T. Lux, M. Marsili and M. Potters.

\end{document}